\documentclass[10pt]{elsarticle}
\usepackage{epstopdf}
\usepackage{graphicx}
\usepackage{amsmath}
\usepackage{url}
\usepackage{color}
\usepackage{wrapfig}
\usepackage{soul}

\journal{Curr. Op. in Neurobiology}

\begin{document}

\begin{frontmatter}



\title{On simplicity and complexity in the brave new world of large-scale neuroscience}


\author[pg]{Peiran Gao}
\ead{prgao@stanford.edu}
\address[pg]{Department of Bioengineering, Stanford University, Stanford, CA 94305}
\author[sg]{Surya Ganguli}
\ead{sganguli@stanford.edu}
\address[sg]{Department of Applied Physics, Stanford University, Stanford, CA 94305}

\begin{abstract}

Technological advances have dramatically expanded our ability to probe multi-neuronal dynamics and connectivity in the brain.  However, our ability to extract a simple conceptual understanding from complex data is increasingly hampered by the lack of theoretically principled data analytic procedures, as well as theoretical frameworks for how circuit connectivity and dynamics can conspire to generate emergent behavioral and cognitive functions.   We review and outline potential avenues for progress, including new theories of high dimensional data analysis, the need to analyze complex artificial networks, and methods for analyzing entire spaces of circuit models, rather than one model at a time.  Such interplay between experiments, data analysis and theory will be indispensable in catalyzing conceptual advances in the age of large-scale neuroscience.

\end{abstract}

\end{frontmatter}





\vspace{\baselineskip}

\par \noindent ``{\it Things should be as simple as possible, but not simpler.}" 
\par \indent -Albert Einstein.

\section*{Introduction}

\par Experimental neuroscience is entering a golden age marked by the advent of remarkable new methods enabling us to record ever increasing numbers of neurons~\cite{stevenson2011,robinson2011,ahrensbrain-wide2012,schrodel2013,ziv2013,prevedel2014}, and measure brain connectivity at various levels of resolution~\cite{micheva2007,wickersham2007,li2010,ragan2012,chung2013,takemura2013,pestilli2014,oh2014}, sometimes measuring both connectivity and dynamics in the same set of neurons~\cite{bock2011,rancz2011}. This recent thrust of technology development is spurred by the hope that an understanding of how the brain gives rise to sensations, actions and thoughts will lurk within the resulting brave new world of complex large-scale data sets.   However, the question of how one can extract a conceptual understanding from data remains a significant challenge for our field.    Major issues involve: (1) What does it even mean to conceptually understand ``how the brain works?'' (2) Are we collecting the right kinds and amounts of data to derive such understanding? (3) Even if we could collect {\it any} kind of detailed measurements about neural structure and function, what theoretical and data analytic procedures would we use to extract conceptual understanding from such measurements?  These are profound questions to which we do not have crisp, detailed answers.   Here we merely present potential routes towards the beginnings of progress on these fronts.

\section*{Understanding as a journey from complexity to simplicity}

\par First, the vague question of ``how the brain works" can be meaningfully reduced to the more precise, and proximally answerable question of how do the connectivity and dynamics of distributed neural circuits give rise to specific behaviors and computations?  But what would a satisfactory answer to this question look like? A detailed, predictive circuit model down to the level of ion-channels and synaptic vesicles within individual neurons, while remarkable, may not yield conceptual understanding in any meaningful human sense.   For example, if simulating this detailed circuit were the {\it only} way we could predict behavior, then we would be loath to say that we {\it understand}  how behavior emerges from the brain.   

\par Instead, a good benchmark for understanding  can be drawn from the physical sciences.  Feynman articulated the idea that we understand a physical theory if we can say something about the solutions to the underlying equations of the theory without actually solving those equations.   For example, we understand aspects of fluid mechanics because we can say many things about specific fluid flows,  without having to numerically solve the Navier-Stokes equations  in every single case.   Similarly,  in neuroscience,  understanding will be found when we have the ability to develop simple coarse-grained models, or better yet a hierarchy of models, at varying levels of biophysical detail, all capable of predicting salient aspects of behavior at varying levels of resolution.  In traversing this hierarchy,  we will obtain  an invaluable understanding of which biophysical details matter, and more importantly, which don't, for any given behavior.   Thus our goal should be to find simplicity amidst complexity,  while of course keeping in mind Einstein's famous dictum quoted above.

\section*{How many neurons are enough: simplicity and complexity in multineuronal dynamics} 

\par What kinds and amounts of data are required to arrive at simple but accurate coarse grained models? In the world of large scale recordings, where we do not have access to simultaneous connectivity information, the focus has been on obtaining a state-space description of the dynamics of neural circuits through various dimensionality reduction methods (see \cite{cunningham2014dimensionality} for a review).   This body of work raises a key conceptual issue permeating much of systems neuroscience, namely,  what precisely can we infer about neural circuit dynamics and its relation to cognition and behavior while measuring only an infinitesimal fraction of behaviorally relevant neurons?   For example, given a doubling time of about 7.4 years \cite{stevenson2011advances} in the number of neurons we can simultaneously measure at single cell, single spike-time resolution, we would have to wait more than 100 years before we can observe $O(10^6-10^9)$ neurons typically present in full mammalian circuits controlling complex behaviors \cite{shepherd2004synaptic}.   Thus, systems neuroscience will remain for the foreseeable future within the vastly undersampled measurement regime, so we need a  {\it theory} of neuronal data analysis in this regime.  Such theory is essential for (1) guiding the biological interpretation of complex multivariate data analytic techniques, (2) efficiently designing future large scale recording experiments, and (3) developing theoretically principled data analysis algorithms appropriate for the degree of subsampling.   

\begin{figure}[h]
      \centering
      \includegraphics[scale=1.0]{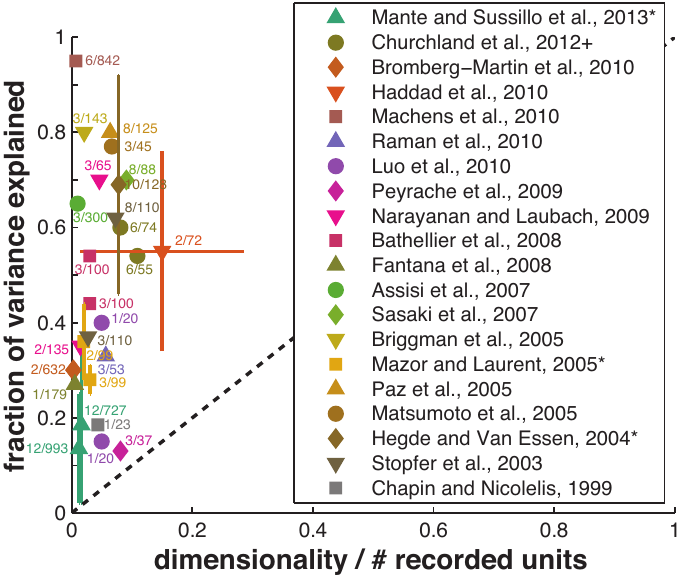}
	\caption{\footnotesize{In many experiments (e.g. in insect~\cite{stopferintensity2003,mazortransient2005,assisiadaptive2007,ramantemporally2010,haddadglobal2010} olfactory systems, mammalian olfactory~\cite{bathellierdynamic2008,haddadglobal2010}, 
prefrontal ~\cite{narayanandelay2009,peyrachereplay2009,machensfunctional2010,wardentask-dependent2010,mantecontext-dependent2013}, motor and premotor,\cite{pazemerging2005,churchlandneural2012}, somatosensory ~\cite{chapinprincipal1999}, visual ~\cite{hegdetemporal2004,matsumotopopulation2005}, hippocampal~\cite{sasakimetastability2007}, and brain stem~\cite{bromberg-martincoding2010} systems) a {\it much} smaller number of dimensions than the number of recorded neurons captures a large amount of variance in neural firing rates.}} 
	\label{fig:literature}
\end{figure}

A clue to the beginnings of this theory lies in an almost universal result occurring across many experiments in which neuroscientists tightly control behavior, record many trials, and obtain trial averaged neuronal firing rate data from hundreds of neurons: in such experiments, the dimensionality (i.e. number of principal components required to explain a fixed percentage of variance) of neural data turns out to be much less than the number of recorded neurons (Fig. \ref{fig:literature}). Moreover, when dimensionality reduction procedures are used to extract neuronal state dynamics,  the resulting low dimensional neural trajectories yield a remarkably insightful dynamical portrait of circuit computation (e.g. ~\cite{mazortransient2005,machensfunctional2010,mantecontext-dependent2013}).

\begin{figure}[h]
      \centering
      \includegraphics[scale=1.0]{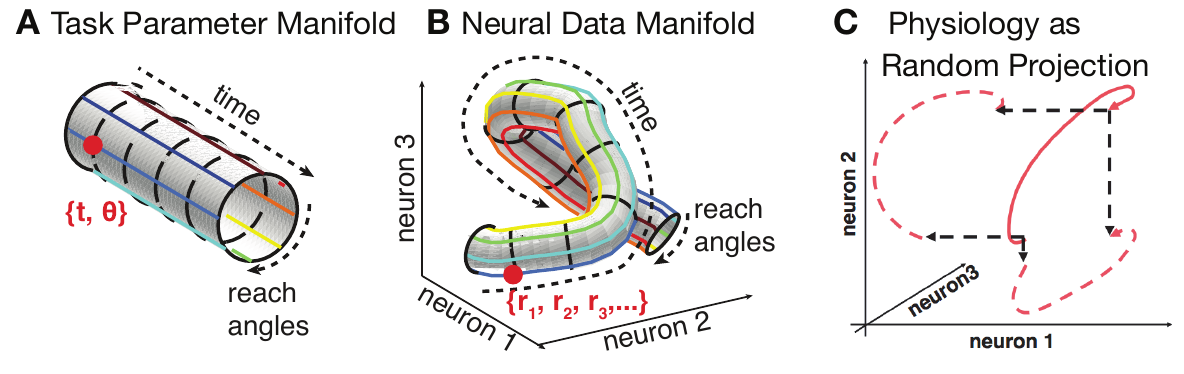}
	\caption{\footnotesize{(A) For a monkey reaching to different directions, the trial averaged behavioral states visited by the arm throughout the experiment are parameterized by a cylinder with two coordinates, reach angle $\theta$, and time into the reach $t$. (B)  Trial averaged neural data is an embedding of the task manifold into firing rate space.  The number of dimensions explored by the neural data manifold is limited by its volume and its curvature (but not the total number of neurons in the motor cortex), with smoother embeddings exploring fewer dimensions.  The NTC is a mathematically precise upper bound on the number of dimensions of the neural data manifold given the volume of the task parameter manifold and a smoothness constraint on the embedding. (C) If the neural data manifold is low dimensional and randomly oriented w.r.t. single neuron axes, then its shadow onto a subset of recorded neurons will preserve its geometric structure.  We have shown, using random projection theory \cite{johnson1984extensions,dasgupta2003elementary, baraniuk2009random} that to preserve neural data manifold geometries with fractional error $\epsilon$, one needs to record $M \geq \frac{1}{\epsilon}K \log (\,\text{NTC}\,)$ neurons.  The figure illustrates a $K=1$ dimensional neural manifold in $N=3$ neurons, and we only record $M=2$ neurons. Thus, fortunately,  the intrinsic complexity of the neural data manifold (small), not the number of neurons in the circuit (large) determines how many neurons we need to record.}} 
	\label{fig:trialavg}
\end{figure}
\par These results raise several profound and timely questions: what is the origin of the underlying simplicity implied by the low dimensionality of neuronal recordings?  How can we trust the dynamical portraits that we extract from so few neurons?  Would the dimensionality increase if we recorded more neurons?  Would the portraits change?  Without an adequate theory, it is impossible to quantitatively answer, or even precisely formulate, these important questions.  We have recently started to develop such a theory \cite{ganguli13dimdyncorr, ganguli13dimdyn}.  Central to this theory is the mathematically well-defined notion of neuronal task complexity (NTC).  Intuitively, the NTC measures the volume of the manifold of task parameters (see Fig. \ref{fig:trialavg}A for the special cases of simple reaches) measured in units of the neuronal population autocorrelation scale across each task parameter.  Thus the NTC in essence measures how many neuronal activity patterns could possibly appear during the course of an experiment given that task parameters have a limited extent and neuronal activity patterns vary smoothly across task parameters (Fig. \ref{fig:trialavg}B). With the mathematical definition of the NTC in hand, we derive that (1) the dimensionality of neuronal data is upper bounded by the NTC, and (2) if the neural data manifold is sufficiently randomly oriented, we can accurately recover dynamical portraits when the number of observed neurons is proportional to the log of the NTC (Fig. \ref{fig:trialavg}C).
     
\par These theorems have significant implications for the interpretation and design of  large-scale experiments.  First, it is likely that in a wide variety of experiments, the origin of low dimensionality is due to a small NTC, a hypothesis that we have verified in recordings from the motor and premotor cortices of monkeys performing a simple 8 direction reach task \cite{byron2007mixture}.   In any such scenario, simply increasing the number of recorded neurons, without a concomitant increase in task complexity will not lead to richer, higher dimensional datasets - indeed data dimensionality will be independent of the number of recorded neurons.  Moreover, we confirmed in motor cortical data our theoretically predicted result that the number of recorded neurons should be proportional to the logarithm of the NTC to accurately recover dynamical portraits of neural state trajectories.  This is excellent news:  while we must make tasks more complex to obtain richer, more insightful datasets, we need not record from many more neurons within a brain region to accurately recover its internal state-space dynamics.

\section*{Towards a theory of single trial data analysis}

\begin{figure}[h]
      \centering
      \includegraphics[width=\textwidth]{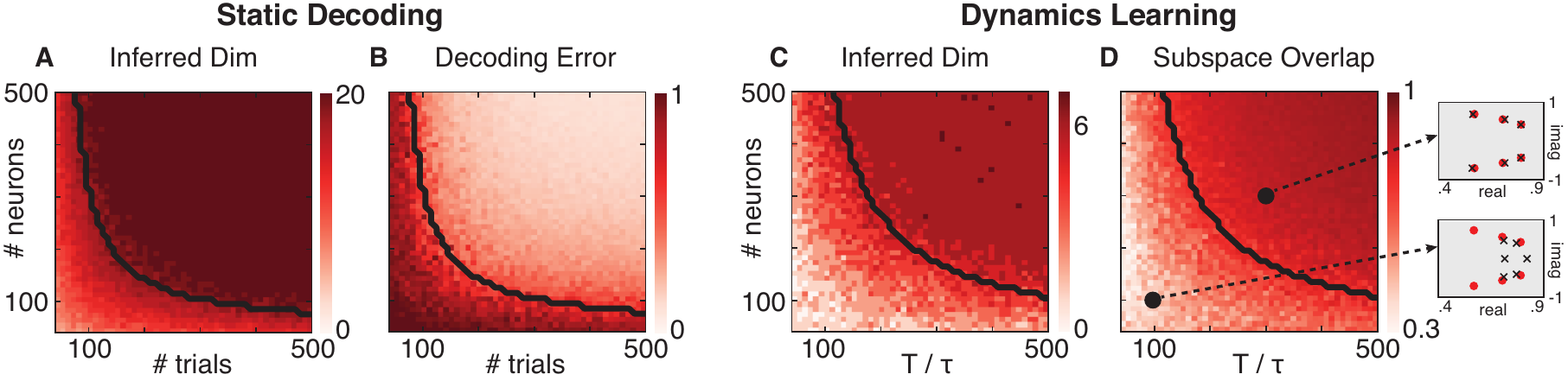}
	\caption{\footnotesize{(A) and (B) The inferred dimensionality and held-out single trial decoding error as a function of $P$ simulated training examples (single trials) and $M$ recorded neurons in a situation where stimuli are encoded in a $K=20$ dimensional subspace in a network of $N=5000$ neurons, with SNR=5.   Inference was performed using low rank matrix denoising \cite{Gavish_Donoho_2014}, and our new analysis of this algorithm reveals a sufficient condition for accurate inference, {\tiny{$\text{SNR}\sqrt{\frac{M}{P}}(\sqrt{P} - \sqrt{K})^2\left(\sqrt{\frac{N - K}{N}} - \sqrt{\frac{K}{M}}\sqrt{\frac{N - M}{N}}\right)^2 \geq K$}}.  The black curve in (A) and (B) reflects the theoretically predicted phase boundary in the $P$, $M$ plane separating accurate from inaccurate inference.  This expression simplifies in the experimentally relevant regime {\tiny{$K, M \ll N$ and $K \ll M, P$ to $\text{SNR}\sqrt{MP} >  K$}}.  (C) and (D) Learning the dimensionality and dynamics, via subspace identification \cite{lennart1999system} of a linear neural network of size $N=5000$ from spontaneous noise driven activity.  The low-rank connectivity of the network forces the system to lie in a $K=6$ dimensional subspace.  Performance is measured as a function of the number of recorded neurons $M$ and recording time $T$.  By combining and extending time series random matrix theory \cite{Yao_2012}, low-rank perturbation theory \cite{Benaych-Georges_Nadakuditi_2012}, and noncommutative probability theory, \cite{Nica_Speicher_1996}, we have derived a theoretically predicted phase boundary (black curve in (C) and(D)), that matches simulations.  In (D), left, the subspace overlap is the correlation between the inferred subspace and the true subspace, with $1$ being perfect correlation, or overlap. In (D), on the right, dynamics (eigenvalues) are correctly predicted only on the right side of the boundary (red dots are true eigenvalues, blue crosses are estimated eigenvalues).}}
	\vspace{-0.75em}
	\label{fig:singletrial}
\end{figure}

The above work suggests that the proximal route for progress lies not in recording more neurons alone, but in designing more complex tasks and stimuli.   However,  with such increased complexity, the same behavioral state or stimulus may rarely be revisited, precluding the possibility of trial averaging as a method for data analysis. Therefore it is essential to extend our theory to the case of single trial analysis.  A simple formulation of the problem is as follows: suppose we have a $K$ dimensional manifold of behavioral states (or stimuli), where $K$ is not necessarily known, and the animal explores $P$  states in succession.  The behavior is controlled by a circuit with $N$ neurons but we measure only $M$ of them. Furthermore, each neuron is noisy with a finite SNR, reflecting single trial variability.  For what values of $M$, $N$, $P$, $K$, and the SNR can we accurately (1) estimate the dimensionality $K$ of neural data, and (2) accurately decode behavior on single trials?  We have solved this problem analytically in the case in which noisy neural activity patterns reflecting $P$ discrete stimuli lie near a  $K$ dimensional subspace (Fig. \ref{fig:singletrial}AB).  We find, roughly that the relations $M,P > K$ and $\text{SNR}\sqrt{MP} >  K$ are sufficient. Thus, it is an intrinsic measure of neural complexity $K$, and not the total number of neurons $N$, that sets a lower bound on how many neurons $M$ and stimuli $P$ we must observe at a given SNR for accurate single trial analyses.  Moreover, we have generalized this analysis to learning dynamical systems (Fig. \ref{fig:singletrial}CD). 

\par Both our preliminary analyses reveal the existence of phase transitions in performance as a function of (1) the number of recorded neurons, and (2) the amount of recording time, stimuli, or behavioral states. Only on the correct side of the phase boundary are accurate dimensionality, dynamics estimation and single trial decoding possible.  Such phase transitions are often found in many high dimensional data analysis problems \cite{amelunxen2014living}, for example in compressed sensing \cite{donoho2009message, ganguli2010statistical} and matrix recovery \cite{donoho2013phase}.  They reveal the beautiful fact that we can recover a lot of information about large objects (vectors or matrices) using surprisingly few measurements when we have seemingly weak prior information, like sparsity, or low-rank structure (see \cite{ganguli2012annrevs,advani2013statistical} for reviews in a neuroscience context).  Moreover, in Fig. \ref{fig:singletrial}, we see that we can move along the phase boundary by trading off number of recorded neurons with recording time. 

\par Thus, to guide future large-scale experimental design, it will be exceedingly important to determine the position of these phase boundaries under increasingly realistic biological assumptions, for example exploring the roles of spiking variability, noise correlations, sparsity, cell types, and network connectivity constraints, and how they impact our ability to uncover true network dynamics and single trial decodes in the face of subsampling.  In essence, we need to develop a Rosetta stone connecting biophysical network dynamics to statistics.  This dictionary will teach us when and how the learned parameters of statistical models fit to a subset of recorded neurons ultimately encode the collective dynamics of the much larger, unobserved neural circuit containing them - an absolutely fundamental question in neuroscience.

\section*{Understanding complex networks with complete information}


\par As we increasingly obtain information about both the connectivity and dynamics of neural circuits, we have to ask ourselves how should we use this information?  As a way to sharpen our ideas, it can be useful to engage in a thought experiment in which experimental neuroscience eventually achieves complete success, in enabling us to measure detailed connectivity, dynamics and plasticity in full neural sub-circuits during behavior. How then would we extract understanding from such rich data?  Moreover, could we arrive at this same understanding without collecting all the data, perhaps even only collecting data within reach in the near future? To address this thought experiment, it is useful to turn to advances in computer science, where deep or recurrent neural networks, consisting of multiple layers of cascaded nonlinearities,  have made a resurgence as the method of choice for solving a range of difficult computational problems.  Indeed, deep learning (see \cite{bengio2013representation, bengio2009learning, schmidhuber2015deep} for reviews)  has led to advances in object detection \cite{krizhevsky2012imagenet, szegedy2014going}, face recognition \cite{sun2014deep, taigman2014deepface}, speech recognition \cite{hannun2014deep}, language translation \cite{sutskever2014sequence}, genomics \cite{xiong2015human}, microscopy \cite{ciresan2012deep}, and even modeling biological neural responses \cite{serre2007feedforward, yamins2014performance, cadieu2014deep, agrawal2014pixels}.  
 
 \par Each of these networks can solve a complex computational problem.  Moreover, we know the full network connectivity, the dynamics of every single neuron, the plasticity rule used to train the network, and indeed the entire developmental experience of the network, in terms of its exposure to training stimuli. Virtually any experiment we wish to do on these networks, we can do.  Yet a meaningful understanding of how these networks work still eludes us, as well has what a suitable benchmark for such understanding would be.  Following Feynman's guideline for understanding physical theories, can we say something about the behavior of deep or recurrent artificial neural networks {\it without} actually simulating them in detail? More importantly, could we arriving at these statements of understanding without measuring every detail of the network, and what are the minimal set of measurements we would need? We do not believe that understanding these networks will directly inform us about how much more complex biological neural networks operate.  However, even in an artificial setting, directly confronting the question of what it means to understand how complex distributed circuits compute, and what kinds of experiments and data analytic procedures we would need to arrive at this understanding, could have a dramatic impact on the questions we ask, experiments we design, and the data analysis we do, in the pursuit of this same understanding in biological neural circuits.  
 
 \par The {\it theory} of deep learning is still in its infancy, but some examples of general statements one can make about deep circuits without actually simulating them include how their functional complexity scales with depth \cite{bianchini2014complexity, pascanu2014number}, how their synaptic weights, over time, acquire statistical structures in inputs \cite{ganguli13exactsol, ganguli13learning}, and how their learning dynamics is dominated by saddle points, not local minima \cite{ganguli13exactsol, dauphin2014identifying}.   However, much more work at the intersection of experimental and theoretical neuroscience and machine learning will be required before we can address the intriguing thought experiment of what we should do if we could measure anything we wanted.    
 
\section*{Understanding not a single model, but the space of all possible models}

An even higher level of understanding is achieved when we develop not just a single model that explains a data set,  but rather understand the space of all possible models consistent with the data.   Such an understanding can place existing biological systems within their evolutionary context,  leading to insights about why they are structured the way they are,  and can reveal general principles that transcend any particular model.  Inspiring examples for neuroscientists can be found not only within neuroscience, but also in allied fields.   For example \cite{li2004yeast} derived a single boolean network model of the yeast cell-cycle control network, while \cite{Lau:2007:Phys-Rev-E-Stat-Nonlin-Soft-Matter-Phys:17677098} developed methods to count and sample from the space of all networks that realize the yeast cell-cycle.  This revealed an astronomical number of possible networks consistent with the data, but only $3\%$ of these networks were more robust than the one chosen by nature, revealing potential evolutionary pressure towards robustness.  In protein folding, theoretical work \cite{li1996emergence} analyzed, in toy models, the space of all possible amino acid sequences that give rise to a given fold; the number of such sequences is the designability of the fold.  Theory revealed that typical folds with shapes similar to those occurring in nature are highly designable, and therefore more easily found by evolution.  Moreover, designable folds are thermodynamically stable \cite{wingreen2004designability} and atypical in shape \cite{li1998protein}, revealing general principles relating sequence to structure.  In the realm of short-term sequence memory, the idea of liquid state machines \cite{maass2002real,Jaeger2001} posited that generic neural circuits could convert temporal information into instantaneous spatial patterns of activity, but theoretical work \cite{white2004short,ganguli2008memory,ganguli2010short} revealed general principles relating circuit connectivity to memory, highlighting the role of non-normal and orthogonal network connectivities in achieving robust sequence memory.  In the realm of long-term memory, seminal work revealed that it is essential to treat synapses as entire dynamical systems in their own right, exhibiting a particular synaptic model \cite{fusi2005cascade}, while further theoretical work \cite{ganguli14memfrontier} analyzed the space of all possible synaptic dynamical systems, revealing general principles relating synaptic structure to function.  Furthermore conductance based models of central pattern generators revealed that highly disparate conductance levels can yield similar behavior \cite{prinz2004similar}, suggesting that observed correlations in conductance levels across animals \cite{schulz2007quantitative} could reflect a signature of homeostatic design \cite{o2013correlations}.   

\par These examples all show that studying the space of models consistent with a given data set or behavior can greatly expand our conceptual understanding.  Further work along these lines within the context of neuronal networks are likely to yield important insights.  For example, suppose we could understand  the space of all possible deep or recurrent neural networks that solve a given  computational task.  Which observable aspects of the connectivity and dynamics are universal across this space,  and which are highly variable across individual networks?   Are the former observables the ones we should focus on measuring in real biological circuits solving the same task?  Are the latter observables less relevant and more indicative of historical accidents over the time course of learning?   

 \par In summary, there are great challenges and opportunities for generating advances in high dimensional data analysis and neuronal circuit theory that can aid in not only responding to the need to interpret existing complex data, but also in driving the questions we ask, and the design of large-scale experiments we do to answer these questions.  Such advances in theory and data analysis will be required to transport us from the ``brave new world, that has such [complex technology] in't'' \cite{shakespeare2001tempest} and deliver us to the promised land of conceptual understanding.  

\section*{Acknolwedgements}

The authors thank Ben Poole, Zayd Enam, Niru Maheswaranathan, and other members of the Neural Dynamics and Computation Lab at Stanford for interesting discussions.  We thank Eric Trautmann and Krishna Shenoy who collaborated with us on the theory of trial averaged dimensionality reduction.  We also thank the ONR and the Burroughs-Wellcome, Sloan, and McDonnell foundations, and the Stanford Center for Mind Brain and Computation for funding. 

\bibliographystyle{model3-num-names}

\begin{thebibliography}{90}
\providecommand{\natexlab}[1]{#1}
\providecommand{\url}[1]{\texttt{#1}}
\providecommand{\urlprefix}{URL }
\expandafter\ifx\csname urlstyle\endcsname\relax
  \providecommand{\doi}[1]{doi:\discretionary{}{}{}#1}\else
  \providecommand{\doi}{doi:\discretionary{}{}{}\begingroup
  \urlstyle{rm}\Url}\fi
\providecommand{\eprint}[2][]{\url{#2}}
\providecommand{\BIBand}{and}
\providecommand{\bibinfo}[2]{#2}
\ifx\xfnm\undefined \def\xfnm[#1]{\unskip,\space#1}\fi
\bibitem[{Stevenson and Kording(2011{\natexlab{a}})}]{stevenson2011}
\bibinfo{author}{Stevenson\xfnm[ I.H.]}, \bibinfo{author}{Kording\xfnm[ K.P.]}.
\newblock \bibinfo{title}{How advances in neural recording affect data
  analysis}.
\newblock \bibinfo{journal}{Nat Neurosci}
  \bibinfo{year}{2011}{\natexlab{a}};\bibinfo{volume}{14}(\bibinfo{number}{2}):\bibinfo{pages}{139--142}.
\bibitem[{Robinson et~al.(2012)Robinson, Jorgolli, Shalek, Yoon, Gertner and
  Park}]{robinson2011}
\bibinfo{author}{Robinson\xfnm[ J.T.]}, \bibinfo{author}{Jorgolli\xfnm[ M.]},
  \bibinfo{author}{Shalek\xfnm[ A.K.]}, \bibinfo{author}{Yoon\xfnm[ M.H.]},
  \bibinfo{author}{Gertner\xfnm[ R.S.]}, \bibinfo{author}{Park\xfnm[ H.]}.
\newblock \bibinfo{title}{Vertical nanowire electrode arrays as a scalable
  platform for intracellular interfacing to neuronal circuits}.
\newblock \bibinfo{journal}{Nat Nano}
  \bibinfo{year}{2012};\bibinfo{volume}{7}(\bibinfo{number}{3}):\bibinfo{pages}{180--184}.
\bibitem[{Ahrens et~al.(2012)Ahrens, Li, Orger, Robson, Schier, Engert
  et~al.}]{ahrensbrain-wide2012}
\bibinfo{author}{Ahrens\xfnm[ M.B.]}, \bibinfo{author}{Li\xfnm[ J.M.]},
  \bibinfo{author}{Orger\xfnm[ M.B.]}, \bibinfo{author}{Robson\xfnm[ D.N.]},
  \bibinfo{author}{Schier\xfnm[ A.F.]}, \bibinfo{author}{Engert\xfnm[ F.]},
  et~al.
\newblock \bibinfo{title}{Brain-wide neuronal dynamics during motor adaptation
  in zebrafish.}
\newblock \bibinfo{journal}{Nature}
  \bibinfo{year}{2012};\bibinfo{volume}{485}(\bibinfo{number}{7399}):\bibinfo{pages}{471--477}.
\bibitem[{Schrodel et~al.(2013)Schrodel, Prevedel, Aumayr, Zimmer and
  Vaziri}]{schrodel2013}
\bibinfo{author}{Schrodel\xfnm[ T.]}, \bibinfo{author}{Prevedel\xfnm[ R.]},
  \bibinfo{author}{Aumayr\xfnm[ K.]}, \bibinfo{author}{Zimmer\xfnm[ M.]},
  \bibinfo{author}{Vaziri\xfnm[ A.]}.
\newblock \bibinfo{title}{Brain-wide 3d imaging of neuronal activity in
  caenorhabditis elegans with sculpted light}.
\newblock \bibinfo{journal}{Nat Meth}
  \bibinfo{year}{2013};\bibinfo{volume}{10}(\bibinfo{number}{10}):\bibinfo{pages}{1013--1020}.
\bibitem[{Ziv et~al.(2013)Ziv, Burns, Cocker, Hamel, Ghosh, Kitch
  et~al.}]{ziv2013}
\bibinfo{author}{Ziv\xfnm[ Y.]}, \bibinfo{author}{Burns\xfnm[ L.D.]},
  \bibinfo{author}{Cocker\xfnm[ E.D.]}, \bibinfo{author}{Hamel\xfnm[ E.O.]},
  \bibinfo{author}{Ghosh\xfnm[ K.K.]}, \bibinfo{author}{Kitch\xfnm[ L.J.]},
  et~al.
\newblock \bibinfo{title}{Long-term dynamics of ca1 hippocampal place codes}.
\newblock \bibinfo{journal}{Nat Neurosci}
  \bibinfo{year}{2013};\bibinfo{volume}{16}(\bibinfo{number}{3}):\bibinfo{pages}{264--266}.
\bibitem[{Prevedel et~al.(2014)Prevedel, Yoon, Hoffmann and Pak}]{prevedel2014}
\newblock {*}
\bibinfo{author}{Prevedel\xfnm[ R.]}, \bibinfo{author}{Yoon\xfnm[ Y.]},
  \bibinfo{author}{Hoffmann\xfnm[ M.]}, \bibinfo{author}{Pak\xfnm[ N.]}.
\newblock \bibinfo{title}{Simultaneous whole-animal 3d imaging of neuronal
  activity using light-field microscopy}.
\newblock \bibinfo{journal}{Nature Methods} \bibinfo{year}{2014}.\\
\newblock {\textbf{Technological advances have dramatically expanded our ability to probe multi-neuronal dynamics and connectivity in the brain. However, our ability to extract a simple conceptual understanding from complex data is increasingly hampered by the lack of theoretically principled data analytic procedures, as well as the- oretical frameworks for how circuit connectivity and dynamics can conspire to generate emergent behavioral and cognitive functions. We review and outline potential avenues for progress, including new theories of high dimensional data analysis, the need to analyze complex artificial networks, and methods for analyzing entire spaces of circuit models, rather than one model at a time. Such interplay between experiments, data analysis and theory will be indispensable in catalyzing conceptual advances in the age of large-scale neuroscience.}}
\bibitem[{Micheva and Smith(2007)}]{micheva2007}
\bibinfo{author}{Micheva\xfnm[ K.]}, \bibinfo{author}{Smith\xfnm[ S.]}.
\newblock \bibinfo{title}{Array tomography: a new tool for imaging the
  molecular architecture and ultrastructure of neural circuits}.
\newblock \bibinfo{journal}{Neuron}
  \bibinfo{year}{2007};\bibinfo{volume}{55}:\bibinfo{pages}{25--36}.
\bibitem[{Wickersham et~al.(2007)Wickersham, Lyon, Barnard, Mori, Finke,
  Conzelmann et~al.}]{wickersham2007}
\bibinfo{author}{Wickersham\xfnm[ I.R.]}, \bibinfo{author}{Lyon\xfnm[ D.C.]},
  \bibinfo{author}{Barnard\xfnm[ R.J.]}, \bibinfo{author}{Mori\xfnm[ T.]},
  \bibinfo{author}{Finke\xfnm[ S.]}, \bibinfo{author}{Conzelmann\xfnm[ K.K.]},
  et~al.
\newblock \bibinfo{title}{Monosynaptic restriction of transsynaptic tracing
  from single, genetically targeted neurons}.
\newblock \bibinfo{journal}{Neuron}
  \bibinfo{year}{2007};\bibinfo{volume}{53}(\bibinfo{number}{5}):\bibinfo{pages}{639
  -- 647}.
\bibitem[{Li et~al.(2010)Li, Gong, Zhang, Wang, Yan, Wu et~al.}]{li2010}
\bibinfo{author}{Li\xfnm[ A.]}, \bibinfo{author}{Gong\xfnm[ H.]},
  \bibinfo{author}{Zhang\xfnm[ B.]}, \bibinfo{author}{Wang\xfnm[ Q.]},
  \bibinfo{author}{Yan\xfnm[ C.]}, \bibinfo{author}{Wu\xfnm[ J.]}, et~al.
\newblock \bibinfo{title}{Micro-optical sectioning tomography to obtain a
  high-resolution atlas of the mouse brain}.
\newblock \bibinfo{journal}{Science}
  \bibinfo{year}{2010};\bibinfo{volume}{330}(\bibinfo{number}{6009}):\bibinfo{pages}{1404--1408}.
\bibitem[{Ragan et~al.(2012)Ragan, Kadiri, Venkataraju, Bahlmann, Sutin,
  Taranda et~al.}]{ragan2012}
\bibinfo{author}{Ragan\xfnm[ T.]}, \bibinfo{author}{Kadiri\xfnm[ L.R.]},
  \bibinfo{author}{Venkataraju\xfnm[ K.U.]}, \bibinfo{author}{Bahlmann\xfnm[
  K.]}, \bibinfo{author}{Sutin\xfnm[ J.]}, \bibinfo{author}{Taranda\xfnm[ J.]},
  et~al.
\newblock \bibinfo{title}{Serial two-photon tomography for automated ex vivo
  mouse brain imaging}.
\newblock \bibinfo{journal}{Nature Methods}
  \bibinfo{year}{2012};\bibinfo{volume}{9}(\bibinfo{number}{3}):\bibinfo{pages}{255--258}.
\bibitem[{Chung and Deisseroth(2013)}]{chung2013}
\newblock{*}
\bibinfo{author}{Chung\xfnm[ K.]}, \bibinfo{author}{Deisseroth\xfnm[ K.]}.
\newblock \bibinfo{title}{Clarity for mapping the nervous system}.
\newblock \bibinfo{journal}{Nat Meth}
  \bibinfo{year}{2013};\bibinfo{volume}{10}(\bibinfo{number}{6}):\bibinfo{pages}{508--513}.\\
\newblock {\textbf{Obtains global brain connectivity information through optical microscopy in brains that have been made transparent through the removal of lipids.}}
\bibitem[{Takemura et~al.(2013)Takemura, Bharioke, Lu, Nern, Vitaladevuni,
  Rivlin et~al.}]{takemura2013}
\bibinfo{author}{Takemura\xfnm[ S.y.]}, \bibinfo{author}{Bharioke\xfnm[ A.]},
  \bibinfo{author}{Lu\xfnm[ Z.]}, \bibinfo{author}{Nern\xfnm[ A.]},
  \bibinfo{author}{Vitaladevuni\xfnm[ S.]}, \bibinfo{author}{Rivlin\xfnm[
  P.K.]}, et~al.
\newblock \bibinfo{title}{A visual motion detection circuit suggested by
  drosophila connectomics}.
\newblock \bibinfo{journal}{Nature}
  \bibinfo{year}{2013};\bibinfo{volume}{500}(\bibinfo{number}{7461}):\bibinfo{pages}{175--181}.
\bibitem[{Pestilli et~al.(2014)Pestilli, Yeatman, Rokem, Kay and
  Wandell}]{pestilli2014}
\bibinfo{author}{Pestilli\xfnm[ F.]}, \bibinfo{author}{Yeatman\xfnm[ J.D.]},
  \bibinfo{author}{Rokem\xfnm[ A.]}, \bibinfo{author}{Kay\xfnm[ K.N.]},
  \bibinfo{author}{Wandell\xfnm[ B.A.]}.
\newblock \bibinfo{title}{Evaluation and statistical inference for human
  connectomes}.
\newblock \bibinfo{journal}{Nat Meth}
  \bibinfo{year}{2014};\bibinfo{volume}{11}(\bibinfo{number}{10}):\bibinfo{pages}{1058--1063}.
\bibitem[{Oh et~al.(2014)Oh, Harris, Ng, Winslow, Cain, Mihalas
  et~al.}]{oh2014}
\bibinfo{author}{Oh\xfnm[ S.W.]}, \bibinfo{author}{Harris\xfnm[ J.A.]},
  \bibinfo{author}{Ng\xfnm[ L.]}, \bibinfo{author}{Winslow\xfnm[ B.]},
  \bibinfo{author}{Cain\xfnm[ N.]}, \bibinfo{author}{Mihalas\xfnm[ S.]}, et~al.
\newblock \bibinfo{title}{A mesoscale connectome of the mouse brain}.
\newblock \bibinfo{journal}{Nature}
  \bibinfo{year}{2014};\bibinfo{volume}{508}(\bibinfo{number}{7495}):\bibinfo{pages}{207--214}.
\bibitem[{Bock et~al.(2011)Bock, Lee, Kerlin, Andermann, Hood, Wetzel
  et~al.}]{bock2011}
\newblock {*}
\bibinfo{author}{Bock\xfnm[ D.D.]}, \bibinfo{author}{Lee\xfnm[ W.C.A.]},
  \bibinfo{author}{Kerlin\xfnm[ A.M.]}, \bibinfo{author}{Andermann\xfnm[
  M.L.]}, \bibinfo{author}{Hood\xfnm[ G.]}, \bibinfo{author}{Wetzel\xfnm[
  A.W.]}, et~al.
\newblock \bibinfo{title}{Network anatomy and in vivo physiology of visual
  cortical neurons}.
\newblock \bibinfo{journal}{Nature}
  \bibinfo{year}{2011};\bibinfo{volume}{471}(\bibinfo{number}{7337}):\bibinfo{pages}{177--182}. \\
\newblock{\textbf{Obtains simultaneous functional and anatomical information about the same set of neurons by combining optical imaging with EM microscopy.}}
\bibitem[{Rancz et~al.(2011)Rancz, Franks, Schwarz, Pichler, Schaefer and
  Margrie}]{rancz2011}
\bibinfo{author}{Rancz\xfnm[ E.A.]}, \bibinfo{author}{Franks\xfnm[ K.M.]},
  \bibinfo{author}{Schwarz\xfnm[ M.K.]}, \bibinfo{author}{Pichler\xfnm[ B.]},
  \bibinfo{author}{Schaefer\xfnm[ A.T.]}, \bibinfo{author}{Margrie\xfnm[
  T.W.]}.
\newblock \bibinfo{title}{Transfection via whole-cell recording in vivo:
  bridging single-cell physiology, genetics and connectomics}.
\newblock \bibinfo{journal}{Nat Neurosci}
  \bibinfo{year}{2011};\bibinfo{volume}{14}(\bibinfo{number}{4}):\bibinfo{pages}{527--532}.
\bibitem[{Cunningham and Byron(2014)}]{cunningham2014dimensionality}
\bibinfo{author}{Cunningham\xfnm[ J.P.]}, \bibinfo{author}{Byron\xfnm[ M.Y.]}.
\newblock \bibinfo{title}{Dimensionality reduction for large-scale neural
  recordings}.
\newblock \bibinfo{journal}{Nature neuroscience} \bibinfo{year}{2014};.
\bibitem[{Stevenson and Kording(2011{\natexlab{b}})}]{stevenson2011advances}
\bibinfo{author}{Stevenson\xfnm[ I.H.]}, \bibinfo{author}{Kording\xfnm[ K.P.]}.
\newblock \bibinfo{title}{How advances in neural recording affect data
  analysis}.
\newblock \bibinfo{journal}{Nature neuroscience}
  \bibinfo{year}{2011}{\natexlab{b}};\bibinfo{volume}{14}(\bibinfo{number}{2}):\bibinfo{pages}{139--142}.
\bibitem[{Shepherd et~al.(2004)}]{shepherd2004synaptic}
\bibinfo{author}{Shepherd\xfnm[ G.M.]}, et~al.
\newblock \bibinfo{title}{The synaptic organization of the brain};
  vol.~\bibinfo{volume}{3}.
\newblock \bibinfo{publisher}{Oxford University Press New York};
  \bibinfo{year}{2004}.
\bibitem[{Mazor and Laurent(2005)}]{mazortransient2005}
\bibinfo{author}{Mazor\xfnm[ O.]}, \bibinfo{author}{Laurent\xfnm[ G.]}.
\newblock \bibinfo{title}{Transient dynamics versus fixed points in odor
  representations by locust antennal lobe projection neurons.}
\newblock \bibinfo{journal}{Neuron}
  \bibinfo{year}{2005};\bibinfo{volume}{48}(\bibinfo{number}{4}):\bibinfo{pages}{661--673}.
\bibitem[{Machens et~al.(2010)Machens, Romo and Brody}]{machensfunctional2010}
\bibinfo{author}{Machens\xfnm[ C.K.]}, \bibinfo{author}{Romo\xfnm[ R.]},
  \bibinfo{author}{Brody\xfnm[ C.D.]}.
\newblock \bibinfo{title}{Functional, but not anatomical, separation of
  {"}what{"} and {"}when{"} in prefrontal cortex.}
\newblock \bibinfo{journal}{The Journal of neuroscience : the official journal
  of the Society for Neuroscience}
  \bibinfo{year}{2010};\bibinfo{volume}{30}(\bibinfo{number}{1}):\bibinfo{pages}{350--360}.
\bibitem[{Mante et~al.(2013)Mante, Sussillo, Shenoy and
  Newsome}]{mantecontext-dependent2013}
\newblock{*}
\bibinfo{author}{Mante\xfnm[ V.]}, \bibinfo{author}{Sussillo\xfnm[ D.]},
  \bibinfo{author}{Shenoy\xfnm[ K.V.]}, \bibinfo{author}{Newsome\xfnm[ W.T.]}.
\newblock \bibinfo{title}{Context-dependent computation by recurrent dynamics
  in prefrontal cortex.}
\newblock \bibinfo{journal}{Nature} \bibinfo{year}{2013};.\\
\newblock{\textbf{Elucidates a distributed mechanism for contextual gating in decision making by training artificial recurrent networks to solve the same task a monkey does, and finds that the artificial neurons behave like the monkeyÕs prefrontal neurons.}}
\bibitem[{Stopfer et~al.(2003)Stopfer, Jayaraman and
  Laurent}]{stopferintensity2003}
\bibinfo{author}{Stopfer\xfnm[ M.]}, \bibinfo{author}{Jayaraman\xfnm[ V.]},
  \bibinfo{author}{Laurent\xfnm[ G.]}.
\newblock \bibinfo{title}{Intensity versus identity coding in an olfactory
  system.}
\newblock \bibinfo{journal}{Neuron}
  \bibinfo{year}{2003};\bibinfo{volume}{39}(\bibinfo{number}{6}):\bibinfo{pages}{991--991004}.
\bibitem[{Assisi et~al.(2007)Assisi, Stopfer, Laurent and
  Bazhenov}]{assisiadaptive2007}
\bibinfo{author}{Assisi\xfnm[ C.]}, \bibinfo{author}{Stopfer\xfnm[ M.]},
  \bibinfo{author}{Laurent\xfnm[ G.]}, \bibinfo{author}{Bazhenov\xfnm[ M.]}.
\newblock \bibinfo{title}{Adaptive regulation of sparseness by feedforward
  inhibition.}
\newblock \bibinfo{journal}{Nature neuroscience}
  \bibinfo{year}{2007};\bibinfo{volume}{10}(\bibinfo{number}{9}):\bibinfo{pages}{1176--1184}.
\bibitem[{Raman et~al.(2010)Raman, Joseph, Tang and
  Stopfer}]{ramantemporally2010}
\bibinfo{author}{Raman\xfnm[ B.]}, \bibinfo{author}{Joseph\xfnm[ J.]},
  \bibinfo{author}{Tang\xfnm[ J.]}, \bibinfo{author}{Stopfer\xfnm[ M.]}.
\newblock \bibinfo{title}{Temporally diverse firing patterns in olfactory
  receptor neurons underlie spatiotemporal neural codes for odors.}
\newblock \bibinfo{journal}{The Journal of neuroscience : the official journal
  of the Society for Neuroscience}
  \bibinfo{year}{2010};\bibinfo{volume}{30}(\bibinfo{number}{6}):\bibinfo{pages}{1994--2006}.
\bibitem[{Haddad et~al.(2010)Haddad, Weiss, Khan, Nadler, Mandairon, Bensafi
  et~al.}]{haddadglobal2010}
\bibinfo{author}{Haddad\xfnm[ R.]}, \bibinfo{author}{Weiss\xfnm[ T.]},
  \bibinfo{author}{Khan\xfnm[ R.]}, \bibinfo{author}{Nadler\xfnm[ B.]},
  \bibinfo{author}{Mandairon\xfnm[ N.]}, \bibinfo{author}{Bensafi\xfnm[ M.]},
  et~al.
\newblock \bibinfo{title}{Global features of neural activity in the olfactory
  system form a parallel code that predicts olfactory behavior and perception.}
\newblock \bibinfo{journal}{The Journal of neuroscience : the official journal
  of the Society for Neuroscience}
  \bibinfo{year}{2010};\bibinfo{volume}{30}(\bibinfo{number}{27}):\bibinfo{pages}{9017--9026}.
\bibitem[{Bathellier et~al.(2008)Bathellier, Buhl, Accolla and
  Carleton}]{bathellierdynamic2008}
\bibinfo{author}{Bathellier\xfnm[ B.]}, \bibinfo{author}{Buhl\xfnm[ D.L.]},
  \bibinfo{author}{Accolla\xfnm[ R.]}, \bibinfo{author}{Carleton\xfnm[ A.]}.
\newblock \bibinfo{title}{Dynamic ensemble odor coding in the mammalian
  olfactory bulb: sensory information at different timescales.}
\newblock \bibinfo{journal}{Neuron}
  \bibinfo{year}{2008};\bibinfo{volume}{57}(\bibinfo{number}{4}):\bibinfo{pages}{586--598}.
\bibitem[{Narayanan and Laubach(2009)}]{narayanandelay2009}
\bibinfo{author}{Narayanan\xfnm[ N.S.]}, \bibinfo{author}{Laubach\xfnm[ M.]}.
\newblock \bibinfo{title}{Delay activity in rodent frontal cortex during a
  simple reaction time task.}
\newblock \bibinfo{journal}{Journal of neurophysiology}
  \bibinfo{year}{2009};\bibinfo{volume}{101}(\bibinfo{number}{6}):\bibinfo{pages}{2859--2871}.
\bibitem[{Peyrache et~al.(2009)Peyrache, Khamassi, Benchenane, Wiener and
  Battaglia}]{peyrachereplay2009}
\bibinfo{author}{Peyrache\xfnm[ A.]}, \bibinfo{author}{Khamassi\xfnm[ M.]},
  \bibinfo{author}{Benchenane\xfnm[ K.]}, \bibinfo{author}{Wiener\xfnm[ S.I.]},
  \bibinfo{author}{Battaglia\xfnm[ F.P.]}.
\newblock \bibinfo{title}{Replay of rule-learning related neural patterns in
  the prefrontal cortex during sleep.}
\newblock \bibinfo{journal}{Nature neuroscience}
  \bibinfo{year}{2009};\bibinfo{volume}{12}(\bibinfo{number}{7}):\bibinfo{pages}{919--926}.
\bibitem[{Warden and Miller(2010)}]{wardentask-dependent2010}
\bibinfo{author}{Warden\xfnm[ M.R.]}, \bibinfo{author}{Miller\xfnm[ E.K.]}.
\newblock \bibinfo{title}{Task-dependent changes in short-term memory in the
  prefrontal cortex.}
\newblock \bibinfo{journal}{The Journal of neuroscience : the official journal
  of the Society for Neuroscience}
  \bibinfo{year}{2010};\bibinfo{volume}{30}(\bibinfo{number}{47}):\bibinfo{pages}{15801--15810}.
\bibitem[{Paz et~al.(2005)Paz, Natan, Boraud, Bergman and
  Vaadia}]{pazemerging2005}
\bibinfo{author}{Paz\xfnm[ R.]}, \bibinfo{author}{Natan\xfnm[ C.]},
  \bibinfo{author}{Boraud\xfnm[ T.]}, \bibinfo{author}{Bergman\xfnm[ H.]},
  \bibinfo{author}{Vaadia\xfnm[ E.]}.
\newblock \bibinfo{title}{Emerging patterns of neuronal responses in
  supplementary and primary motor areas during sensorimotor adaptation.}
\newblock \bibinfo{journal}{The Journal of neuroscience : the official journal
  of the Society for Neuroscience}
  \bibinfo{year}{2005};\bibinfo{volume}{25}(\bibinfo{number}{47}):\bibinfo{pages}{10941--10951}.
\bibitem[{Churchland et~al.(2012)Churchland, Cunningham, Kaufman, Foster,
  Nuyujukian, Ryu et~al.}]{churchlandneural2012}
\bibinfo{author}{Churchland\xfnm[ M.M.]}, \bibinfo{author}{Cunningham\xfnm[
  J.P.]}, \bibinfo{author}{Kaufman\xfnm[ M.T.]}, \bibinfo{author}{Foster\xfnm[
  J.D.]}, \bibinfo{author}{Nuyujukian\xfnm[ P.]}, \bibinfo{author}{Ryu\xfnm[
  S.I.]}, et~al.
\newblock \bibinfo{title}{Neural population dynamics during reaching.}
\newblock \bibinfo{journal}{Nature}
  \bibinfo{year}{2012};\bibinfo{volume}{487}(\bibinfo{number}{7405}):\bibinfo{pages}{51--56}.
\bibitem[{Chapin and Nicolelis(1999)}]{chapinprincipal1999}
\bibinfo{author}{Chapin\xfnm[ J.]}, \bibinfo{author}{Nicolelis\xfnm[ M.]}.
\newblock \bibinfo{title}{Principal component analysis of neuronal ensemble
  activity reveals multidimensional somatosensory representations.}
\newblock \bibinfo{journal}{Journal of neuroscience methods}
  \bibinfo{year}{1999};\bibinfo{volume}{94}(\bibinfo{number}{1}):\bibinfo{pages}{121--140}.
\bibitem[{Hegd\'{e} and Van~Essen(2004)}]{hegdetemporal2004}
\bibinfo{author}{Hegd\'{e}\xfnm[ J.]}, \bibinfo{author}{Van~Essen\xfnm[ D.C.]}.
\newblock \bibinfo{title}{Temporal dynamics of shape analysis in macaque visual
  area v2.}
\newblock \bibinfo{journal}{Journal of neurophysiology}
  \bibinfo{year}{2004};\bibinfo{volume}{92}(\bibinfo{number}{5}):\bibinfo{pages}{3030--3042}.
\bibitem[{Matsumoto et~al.(2005)Matsumoto, Okada, Yasuko, Yamane and
  Kawano}]{matsumotopopulation2005}
\bibinfo{author}{Matsumoto\xfnm[ N.]}, \bibinfo{author}{Okada\xfnm[ M.]},
  \bibinfo{author}{Yasuko\xfnm[ S.]}, \bibinfo{author}{Yamane\xfnm[ S.]},
  \bibinfo{author}{Kawano\xfnm[ K.]}.
\newblock \bibinfo{title}{Population dynamics of face-responsive neurons in the
  inferior temporal cortex.}
\newblock \bibinfo{journal}{Cerebral cortex {(New} York, {NY} : 1991)}
  \bibinfo{year}{2005};\bibinfo{volume}{15}(\bibinfo{number}{8}):\bibinfo{pages}{1103--1112}.
\bibitem[{Sasaki et~al.(2007)Sasaki, Matsuki and
  Ikegaya}]{sasakimetastability2007}
\bibinfo{author}{Sasaki\xfnm[ T.]}, \bibinfo{author}{Matsuki\xfnm[ N.]},
  \bibinfo{author}{Ikegaya\xfnm[ Y.]}.
\newblock \bibinfo{title}{Metastability of active {CA3} networks.}
\newblock \bibinfo{journal}{The Journal of neuroscience : the official journal
  of the Society for Neuroscience}
  \bibinfo{year}{2007};\bibinfo{volume}{27}(\bibinfo{number}{3}):\bibinfo{pages}{517--528}.
\bibitem[{Bromberg-Martin et~al.(2010)Bromberg-Martin, Hikosaka and
  Nakamura}]{bromberg-martincoding2010}
\bibinfo{author}{Bromberg-Martin\xfnm[ E.S.]}, \bibinfo{author}{Hikosaka\xfnm[
  O.]}, \bibinfo{author}{Nakamura\xfnm[ K.]}.
\newblock \bibinfo{title}{Coding of task reward value in the dorsal raphe
  nucleus.}
\newblock \bibinfo{journal}{The Journal of neuroscience : the official journal
  of the Society for Neuroscience}
  \bibinfo{year}{2010};\bibinfo{volume}{30}(\bibinfo{number}{18}):\bibinfo{pages}{6262--6272}.
\bibitem[{Johnson and Lindenstrauss(1984)}]{johnson1984extensions}
\bibinfo{author}{Johnson\xfnm[ W.]}, \bibinfo{author}{Lindenstrauss\xfnm[ J.]}.
\newblock \bibinfo{title}{Extensions of lipschitz mappings into a hilbert
  space}.
\newblock \bibinfo{journal}{Contemporary mathematics}
  \bibinfo{year}{1984};\bibinfo{volume}{26}(\bibinfo{number}{189-206}):\bibinfo{pages}{1--1}.
\bibitem[{Dasgupta and Gupta(2003)}]{dasgupta2003elementary}
\bibinfo{author}{Dasgupta\xfnm[ S.]}, \bibinfo{author}{Gupta\xfnm[ A.]}.
\newblock \bibinfo{title}{An elementary proof of a theorem of johnson and
  lindenstrauss}.
\newblock \bibinfo{journal}{Random Structures \& Algorithms}
  \bibinfo{year}{2003};\bibinfo{volume}{22}(\bibinfo{number}{1}):\bibinfo{pages}{60--65}.
\bibitem[{Baraniuk and Wakin(2009)}]{baraniuk2009random}
\newblock{*}
\bibinfo{author}{Baraniuk\xfnm[ R.]}, \bibinfo{author}{Wakin\xfnm[ M.]}.
\newblock \bibinfo{title}{Random projections of smooth manifolds}.
\newblock \bibinfo{journal}{Foundations of Computational Mathematics}
  \bibinfo{year}{2009};\bibinfo{volume}{9}(\bibinfo{number}{1}):\bibinfo{pages}{51--77}.\\
\newblock{\textbf{Reveals that when data or signals lie on a curved low dimensional manifold embedded in a high dimensional space, remarkably few measurements are required to recover the geometry of the manifold.}}
\bibitem[{Gao et~al.(2014{\natexlab{a}})Gao, Trautmann, Yu, Santhanam, Ryu,
  Shenoy et~al.}]{ganguli13dimdyncorr}
\bibinfo{author}{Gao\xfnm[ P.]}, \bibinfo{author}{Trautmann\xfnm[ E.]},
  \bibinfo{author}{Yu\xfnm[ B.]}, \bibinfo{author}{Santhanam\xfnm[ G.]},
  \bibinfo{author}{Ryu\xfnm[ S.]}, \bibinfo{author}{Shenoy\xfnm[ K.]}, et~al.
\newblock \bibinfo{title}{A theory of neural dimensionality and measurement}.
\newblock In: \bibinfo{booktitle}{Computational and Systems Neuroscience
  Conference (COSYNE)}. \bibinfo{year}{2014}{\natexlab{a}},.
\bibitem[{Gao et~al.(2014{\natexlab{b}})Gao, Trautmann, Yu, Santhanam, Ryu,
  K.Shenoy et~al.}]{ganguli13dimdyn}
\bibinfo{author}{Gao\xfnm[ P.]}, \bibinfo{author}{Trautmann\xfnm[ E.]},
  \bibinfo{author}{Yu\xfnm[ B.]}, \bibinfo{author}{Santhanam\xfnm[ G.]},
  \bibinfo{author}{Ryu\xfnm[ S.]}, \bibinfo{author}{K.Shenoy\xfnm[]}, et~al.
\newblock \bibinfo{title}{A theory of neural dimensionaliy, dynamics and
  measurement}.
\newblock \bibinfo{journal}{article in preparation for Neuron}
  \bibinfo{year}{2014}{\natexlab{b}};.
\bibitem[{Byron et~al.(2007)Byron, Kemere, Santhanam, Afshar, Ryu, Meng
  et~al.}]{byron2007mixture}
\bibinfo{author}{Byron\xfnm[ M.Y.]}, \bibinfo{author}{Kemere\xfnm[ C.]},
  \bibinfo{author}{Santhanam\xfnm[ G.]}, \bibinfo{author}{Afshar\xfnm[ A.]},
  \bibinfo{author}{Ryu\xfnm[ S.I.]}, \bibinfo{author}{Meng\xfnm[ T.H.]}, et~al.
\newblock \bibinfo{title}{Mixture of trajectory models for neural decoding of
  goal-directed movements}.
\newblock \bibinfo{journal}{Journal of neurophysiology}
  \bibinfo{year}{2007};\bibinfo{volume}{97}(\bibinfo{number}{5}):\bibinfo{pages}{3763--3780}.
\bibitem[{Gavish and Donoho(2014)}]{Gavish_Donoho_2014}
\bibinfo{author}{Gavish\xfnm[ M.]}, \bibinfo{author}{Donoho\xfnm[ D.]}.
\newblock \bibinfo{title}{Optimal shrinkage of singular values}.
\newblock \bibinfo{journal}{arXiv preprint arXiv:14057511}
  \bibinfo{year}{2014};.
\bibitem[{Lennart(1999)}]{lennart1999system}
\bibinfo{author}{Lennart\xfnm[ L.]}.
\newblock \bibinfo{title}{System identification: theory for the user}.
\newblock \bibinfo{year}{1999}.
\bibitem[{Yao(2012)}]{Yao_2012}
\newblock{*}
\bibinfo{author}{Yao\xfnm[ J.]}.
\newblock \bibinfo{title}{A note on a marcenko-pasteur type theorem for
  time-series.}
\newblock \bibinfo{journal}{Statistics and Probability Letters}
  \bibinfo{year}{2012};\doi{\bibinfo{doi}{10.1016/j.spl.2011.08.011}}.\\
\newblock{\textbf{Describes the structure of noise in high dimensional linear dynamical systems; this structure partially dictates how much of the system, and for how long, we need to measure to uncover its dynamics.}}
\bibitem[{Benaych-Georges and
  Nadakuditi(2012)}]{Benaych-Georges_Nadakuditi_2012}
\bibinfo{author}{Benaych-Georges\xfnm[ F.]}, \bibinfo{author}{Nadakuditi\xfnm[
  R.]}.
\newblock \bibinfo{title}{The singular values and vectors of low rank
  perturbations of large rectangular random matrices}.
\newblock \bibinfo{journal}{Journal of Multivariate Analysis}
  \bibinfo{year}{2012};\bibinfo{volume}{111}:\bibinfo{pages}{120135}.
\newblock \doi{\bibinfo{doi}{10.1016/j.jmva.2012.04.019}}.
\bibitem[{Nica and Speicher(1996)}]{Nica_Speicher_1996}
\bibinfo{author}{Nica\xfnm[ A.]}, \bibinfo{author}{Speicher\xfnm[ R.]}.
\newblock \bibinfo{title}{On the multiplication of free n-tuples of
  noncommutative random variables}.
\newblock \bibinfo{journal}{American Journal of Mathematics}
  \bibinfo{year}{1996};:\bibinfo{pages}{799--837}\doi{\bibinfo{doi}{10.2307/25098492}}.
\newblock \urlprefix\url{http://www.jstor.org/stable/25098492}.
\bibitem[{Amelunxen et~al.(2014)Amelunxen, Lotz, McCoy and
  Tropp}]{amelunxen2014living}
\newblock{*}
\bibinfo{author}{Amelunxen\xfnm[ D.]}, \bibinfo{author}{Lotz\xfnm[ M.]},
  \bibinfo{author}{McCoy\xfnm[ M.B.]}, \bibinfo{author}{Tropp\xfnm[ J.A.]}.
\newblock \bibinfo{title}{Living on the edge: Phase transitions in convex
  programs with random data}.
\newblock \bibinfo{journal}{Information and Inference}
  \bibinfo{year}{2014};:\bibinfo{pages}{iau005}.\\
\newblock{\textbf{Reveals that a wide variety of data analysis algorithms that correspond to convex optimization problems, have phase transitions in various measures of performance that quantify the success of the algorithm.}}
\bibitem[{Donoho et~al.(2009)Donoho, Maleki and Montanari}]{donoho2009message}
\bibinfo{author}{Donoho\xfnm[ D.]}, \bibinfo{author}{Maleki\xfnm[ A.]},
  \bibinfo{author}{Montanari\xfnm[ A.]}.
\newblock \bibinfo{title}{{Message-passing algorithms for compressed sensing}}.
\newblock \bibinfo{journal}{Proc Natl Acad Sci}
  \bibinfo{year}{2009};\bibinfo{volume}{106}(\bibinfo{number}{45}):\bibinfo{pages}{18914}.
\bibitem[{Ganguli and Sompolinsky(2010{\natexlab{a}})}]{ganguli2010statistical}
\bibinfo{author}{Ganguli\xfnm[ S.]}, \bibinfo{author}{Sompolinsky\xfnm[ H.]}.
\newblock \bibinfo{title}{Statistical mechanics of compressed sensing.}
\newblock \bibinfo{journal}{Phys Rev Lett}
  \bibinfo{year}{2010}{\natexlab{a}};\bibinfo{volume}{104}(\bibinfo{number}{18}):\bibinfo{pages}{188701}.
\bibitem[{Donoho et~al.(2013)Donoho, Gavish and Montanari}]{donoho2013phase}
\bibinfo{author}{Donoho\xfnm[ D.L.]}, \bibinfo{author}{Gavish\xfnm[ M.]},
  \bibinfo{author}{Montanari\xfnm[ A.]}.
\newblock \bibinfo{title}{The phase transition of matrix recovery from gaussian
  measurements matches the minimax mse of matrix denoising}.
\newblock \bibinfo{journal}{Proceedings of the National Academy of Sciences}
  \bibinfo{year}{2013};\bibinfo{volume}{110}(\bibinfo{number}{21}):\bibinfo{pages}{8405--8410}.
\bibitem[{Ganguli and Sompolinsky(2012)}]{ganguli2012annrevs}
\bibinfo{author}{Ganguli\xfnm[ S.]}, \bibinfo{author}{Sompolinsky\xfnm[ H.]}.
\newblock \bibinfo{title}{Compressed sensing, sparsity, and dimensionality in
  neuronal information processing and data analysis}.
\newblock \bibinfo{journal}{Annu Rev Neurosci}
  \bibinfo{year}{2012};\bibinfo{volume}{35}:\bibinfo{pages}{485--508}.
\bibitem[{Advani et~al.(2013)Advani, Lahiri and
  Ganguli}]{advani2013statistical}
\bibinfo{author}{Advani\xfnm[ M.]}, \bibinfo{author}{Lahiri\xfnm[ S.]},
  \bibinfo{author}{Ganguli\xfnm[ S.]}.
\newblock \bibinfo{title}{Statistical mechanics of complex neural systems and
  high dimensional data}.
\newblock \bibinfo{journal}{Journal of Statistical Mechanics: Theory and
  Experiment}
  \bibinfo{year}{2013};\bibinfo{volume}{2013}(\bibinfo{number}{03}):\bibinfo{pages}{P03014}.
\bibitem[{Bengio et~al.(2013)Bengio, Courville and
  Vincent}]{bengio2013representation}
\bibinfo{author}{Bengio\xfnm[ Y.]}, \bibinfo{author}{Courville\xfnm[ A.]},
  \bibinfo{author}{Vincent\xfnm[ P.]}.
\newblock \bibinfo{title}{Representation learning: A review and new
  perspectives}.
\newblock \bibinfo{journal}{Pattern Analysis and Machine Intelligence, IEEE
  Transactions on}
  \bibinfo{year}{2013};\bibinfo{volume}{35}(\bibinfo{number}{8}):\bibinfo{pages}{1798--1828}.
\bibitem[{Bengio(2009)}]{bengio2009learning}
\bibinfo{author}{Bengio\xfnm[ Y.]}.
\newblock \bibinfo{title}{Learning deep architectures for ai}.
\newblock \bibinfo{journal}{Foundations and trends{\textregistered} in Machine
  Learning}
  \bibinfo{year}{2009};\bibinfo{volume}{2}(\bibinfo{number}{1}):\bibinfo{pages}{1--127}.
\bibitem[{Schmidhuber(2015)}]{schmidhuber2015deep}
\bibinfo{author}{Schmidhuber\xfnm[ J.]}.
\newblock \bibinfo{title}{Deep learning in neural networks: An overview}.
\newblock \bibinfo{journal}{Neural Networks}
  \bibinfo{year}{2015};\bibinfo{volume}{61}:\bibinfo{pages}{85--117}.
\bibitem[{Krizhevsky et~al.(2012)Krizhevsky, Sutskever and
  Hinton}]{krizhevsky2012imagenet}
\newblock{*}
\bibinfo{author}{Krizhevsky\xfnm[ A.]}, \bibinfo{author}{Sutskever\xfnm[ I.]},
  \bibinfo{author}{Hinton\xfnm[ G.E.]}.
\newblock \bibinfo{title}{Imagenet classification with deep convolutional
  neural networks}.
\newblock In: \bibinfo{booktitle}{Advances in neural information processing
  systems}. \bibinfo{year}{2012}, p. \bibinfo{pages}{1097--1105}.\\
\newblock{\textbf{Seminal work that yielded substantial performance improvements in visual object recognition through deep neural networks.}}
\bibitem[{Szegedy et~al.(2014)Szegedy, Liu, Jia, Sermanet, Reed, Anguelov
  et~al.}]{szegedy2014going}
\bibinfo{author}{Szegedy\xfnm[ C.]}, \bibinfo{author}{Liu\xfnm[ W.]},
  \bibinfo{author}{Jia\xfnm[ Y.]}, \bibinfo{author}{Sermanet\xfnm[ P.]},
  \bibinfo{author}{Reed\xfnm[ S.]}, \bibinfo{author}{Anguelov\xfnm[ D.]},
  et~al.
\newblock \bibinfo{title}{Going deeper with convolutions}.
\newblock \bibinfo{journal}{arXiv preprint arXiv:14094842}
  \bibinfo{year}{2014};.
\bibitem[{Sun et~al.(2014)Sun, Chen, Wang and Tang}]{sun2014deep}
\bibinfo{author}{Sun\xfnm[ Y.]}, \bibinfo{author}{Chen\xfnm[ Y.]},
  \bibinfo{author}{Wang\xfnm[ X.]}, \bibinfo{author}{Tang\xfnm[ X.]}.
\newblock \bibinfo{title}{Deep learning face representation by joint
  identification-verification}.
\newblock In: \bibinfo{editor}{Ghahramani\xfnm[ Z.]},
  \bibinfo{editor}{Welling\xfnm[ M.]}, \bibinfo{editor}{Cortes\xfnm[ C.]},
  \bibinfo{editor}{Lawrence\xfnm[ N.]}, \bibinfo{editor}{Weinberger\xfnm[ K.]},
  editors. \bibinfo{booktitle}{Advances in Neural Information Processing
  Systems 27}. \bibinfo{publisher}{Curran Associates, Inc.};
  \bibinfo{year}{2014}, p. \bibinfo{pages}{1988--1996}.
\bibitem[{Taigman et~al.(2014)Taigman, Yang, Ranzato and
  Wolf}]{taigman2014deepface}
\bibinfo{author}{Taigman\xfnm[ Y.]}, \bibinfo{author}{Yang\xfnm[ M.]},
  \bibinfo{author}{Ranzato\xfnm[ M.]}, \bibinfo{author}{Wolf\xfnm[ L.]}.
\newblock \bibinfo{title}{Deepface: Closing the gap to human-level performance
  in face verification}.
\newblock In: \bibinfo{booktitle}{Computer Vision and Pattern Recognition
  (CVPR), 2014 IEEE Conference on}. \bibinfo{organization}{IEEE};
  \bibinfo{year}{2014}, p. \bibinfo{pages}{1701--1708}.
\bibitem[{Hannun et~al.(2014)Hannun, Case, Casper, Catanzaro, Diamos, Elsen
  et~al.}]{hannun2014deep}
\bibinfo{author}{Hannun\xfnm[ A.Y.]}, \bibinfo{author}{Case\xfnm[ C.]},
  \bibinfo{author}{Casper\xfnm[ J.]}, \bibinfo{author}{Catanzaro\xfnm[ B.C.]},
  \bibinfo{author}{Diamos\xfnm[ G.]}, \bibinfo{author}{Elsen\xfnm[ E.]}, et~al.
\newblock \bibinfo{title}{Deep speech: Scaling up end-to-end speech
  recognition}.
\newblock \bibinfo{journal}{CoRR}
  \bibinfo{year}{2014};\bibinfo{volume}{abs/1412.5567}.
\bibitem[{Sutskever et~al.(2014)Sutskever, Vinyals and
  Le}]{sutskever2014sequence}
\bibinfo{author}{Sutskever\xfnm[ I.]}, \bibinfo{author}{Vinyals\xfnm[ O.]},
  \bibinfo{author}{Le\xfnm[ Q.V.V.]}.
\newblock \bibinfo{title}{Sequence to sequence learning with neural networks}.
\newblock In: \bibinfo{editor}{Ghahramani\xfnm[ Z.]},
  \bibinfo{editor}{Welling\xfnm[ M.]}, \bibinfo{editor}{Cortes\xfnm[ C.]},
  \bibinfo{editor}{Lawrence\xfnm[ N.]}, \bibinfo{editor}{Weinberger\xfnm[ K.]},
  editors. \bibinfo{booktitle}{Advances in Neural Information Processing
  Systems 27}. \bibinfo{publisher}{Curran Associates, Inc.};
  \bibinfo{year}{2014}, p. \bibinfo{pages}{3104--3112}.
\bibitem[{Xiong et~al.(2015)Xiong, Alipanahi, Lee, Bretschneider, Merico, Yuen
  et~al.}]{xiong2015human}
\bibinfo{author}{Xiong\xfnm[ H.Y.]}, \bibinfo{author}{Alipanahi\xfnm[ B.]},
  \bibinfo{author}{Lee\xfnm[ L.J.]}, \bibinfo{author}{Bretschneider\xfnm[ H.]},
  \bibinfo{author}{Merico\xfnm[ D.]}, \bibinfo{author}{Yuen\xfnm[ R.K.]},
  et~al.
\newblock \bibinfo{title}{The human splicing code reveals new insights into the
  genetic determinants of disease}.
\newblock \bibinfo{journal}{Science}
  \bibinfo{year}{2015};\bibinfo{volume}{347}(\bibinfo{number}{6218}):\bibinfo{pages}{1254806}.
\bibitem[{Ciresan et~al.(2012)Ciresan, Giusti, Gambardella and
  Schmidhuber}]{ciresan2012deep}
\bibinfo{author}{Ciresan\xfnm[ D.]}, \bibinfo{author}{Giusti\xfnm[ A.]},
  \bibinfo{author}{Gambardella\xfnm[ L.M.]}, \bibinfo{author}{Schmidhuber\xfnm[
  J.]}.
\newblock \bibinfo{title}{Deep neural networks segment neuronal membranes in
  electron microscopy images}.
\newblock In: \bibinfo{booktitle}{Advances in neural information processing
  systems}. \bibinfo{year}{2012}, p. \bibinfo{pages}{2843--2851}.
\bibitem[{Serre et~al.(2007)Serre, Oliva and Poggio}]{serre2007feedforward}
\bibinfo{author}{Serre\xfnm[ T.]}, \bibinfo{author}{Oliva\xfnm[ A.]},
  \bibinfo{author}{Poggio\xfnm[ T.]}.
\newblock \bibinfo{title}{A feedforward architecture accounts for rapid
  categorization}.
\newblock \bibinfo{journal}{Proceedings of the National Academy of Sciences}
  \bibinfo{year}{2007};\bibinfo{volume}{104}(\bibinfo{number}{15}):\bibinfo{pages}{6424--6429}.
\bibitem[{Yamins et~al.(2014)Yamins, Hong, Cadieu, Solomon, Seibert and
  DiCarlo}]{yamins2014performance}
\newblock{*}
\bibinfo{author}{Yamins\xfnm[ D.L.]}, \bibinfo{author}{Hong\xfnm[ H.]},
  \bibinfo{author}{Cadieu\xfnm[ C.F.]}, \bibinfo{author}{Solomon\xfnm[ E.A.]},
  \bibinfo{author}{Seibert\xfnm[ D.]}, \bibinfo{author}{DiCarlo\xfnm[ J.J.]}.
\newblock \bibinfo{title}{Performance-optimized hierarchical models predict
  neural responses in higher visual cortex}.
\newblock \bibinfo{journal}{Proceedings of the National Academy of Sciences}
  \bibinfo{year}{2014};:\bibinfo{pages}{201403112}.\\
\newblock{\textbf{Deep neural circuits that were optimized for performance in object categorization contained small collections of neurons in deep layers, whose linear combinations mimicked the responses of neurons in monkey inferotemporal cortex to natural images.}}
\bibitem[{Cadieu et~al.(2014)Cadieu, Hong, Yamins, Pinto, Ardila, Solomon
  et~al.}]{cadieu2014deep}
\bibinfo{author}{Cadieu\xfnm[ C.F.]}, \bibinfo{author}{Hong\xfnm[ H.]},
  \bibinfo{author}{Yamins\xfnm[ D.L.]}, \bibinfo{author}{Pinto\xfnm[ N.]},
  \bibinfo{author}{Ardila\xfnm[ D.]}, \bibinfo{author}{Solomon\xfnm[ E.A.]},
  et~al.
\newblock \bibinfo{title}{Deep neural networks rival the representation of
  primate it cortex for core visual object recognition}.
\newblock \bibinfo{journal}{PLoS computational biology}
  \bibinfo{year}{2014};\bibinfo{volume}{10}(\bibinfo{number}{12}):\bibinfo{pages}{e1003963}.
\bibitem[{Agrawal et~al.(2014)Agrawal, Stansbury, Malik and
  Gallant}]{agrawal2014pixels}
\bibinfo{author}{Agrawal\xfnm[ P.]}, \bibinfo{author}{Stansbury\xfnm[ D.]},
  \bibinfo{author}{Malik\xfnm[ J.]}, \bibinfo{author}{Gallant\xfnm[ J.L.]}.
\newblock \bibinfo{title}{Pixels to voxels: Modeling visual representation in
  the human brain}.
\newblock \bibinfo{journal}{arXiv preprint arXiv:14075104}
  \bibinfo{year}{2014};.
\bibitem[{Bianchini and Scarselli(2014)}]{bianchini2014complexity}
\bibinfo{author}{Bianchini\xfnm[ M.]}, \bibinfo{author}{Scarselli\xfnm[ F.]}.
\newblock \bibinfo{title}{On the complexity of neural network classifiers: A
  comparison between shallow and deep architectures}.
\newblock \bibinfo{journal}{IEEE Transactions on Neural Networks}
  \bibinfo{year}{2014};.
\bibitem[{Pascanu et~al.(2014)Pascanu, Mont{\'u}far and
  Bengio}]{pascanu2014number}
\bibinfo{author}{Pascanu\xfnm[ R.]}, \bibinfo{author}{Mont{\'u}far\xfnm[ G.]},
  \bibinfo{author}{Bengio\xfnm[ Y.]}.
\newblock \bibinfo{title}{On the number of inference regions of deep feed
  forward networks with piece-wise linear activations}.
\newblock \bibinfo{journal}{Internal Conference on Learning Representations
  (ICLR)} \bibinfo{year}{2014};.
\bibitem[{Saxe et~al.(2014)Saxe, McClelland and Ganguli}]{ganguli13exactsol}
\bibinfo{author}{Saxe\xfnm[ A.]}, \bibinfo{author}{McClelland\xfnm[ J.]},
  \bibinfo{author}{Ganguli\xfnm[ S.]}.
\newblock \bibinfo{title}{Exact solutions to the nonlinear dynamics of learning
  in deep linear neural networks}.
\newblock In: \bibinfo{booktitle}{International Conference on Learning
  Representations (ICLR)}. \bibinfo{year}{2014},.
\bibitem[{Saxe et~al.(2013)Saxe, McClelland and Ganguli}]{ganguli13learning}
\bibinfo{author}{Saxe\xfnm[ A.M.]}, \bibinfo{author}{McClelland\xfnm[ J.L.]},
  \bibinfo{author}{Ganguli\xfnm[ S.]}.
\newblock \bibinfo{title}{Learning hierarchical category structure in deep
  neural networks}.
\newblock In: \bibinfo{booktitle}{Proc. of 35th annual Cog. Sci. Society}.
  \bibinfo{year}{2013},.
\bibitem[{Dauphin et~al.(2014)Dauphin, Pascanu, Gulcehre, Cho, Ganguli and
  Bengio}]{dauphin2014identifying}
\bibinfo{author}{Dauphin\xfnm[ Y.N.]}, \bibinfo{author}{Pascanu\xfnm[ R.]},
  \bibinfo{author}{Gulcehre\xfnm[ C.]}, \bibinfo{author}{Cho\xfnm[ K.]},
  \bibinfo{author}{Ganguli\xfnm[ S.]}, \bibinfo{author}{Bengio\xfnm[ Y.]}.
\newblock \bibinfo{title}{Identifying and attacking the saddle point problem in
  high-dimensional non-convex optimization}.
\newblock In: \bibinfo{booktitle}{Advances in Neural Information Processing
  Systems}. \bibinfo{year}{2014}, p. \bibinfo{pages}{2933--2941}.
\bibitem[{Li et~al.(2004)Li, Long, Lu, Ouyang and Tang}]{li2004yeast}
\bibinfo{author}{Li\xfnm[ F.]}, \bibinfo{author}{Long\xfnm[ T.]},
  \bibinfo{author}{Lu\xfnm[ Y.]}, \bibinfo{author}{Ouyang\xfnm[ Q.]},
  \bibinfo{author}{Tang\xfnm[ C.]}.
\newblock \bibinfo{title}{The yeast cell-cycle network is robustly designed}.
\newblock \bibinfo{journal}{Proceedings of the National Academy of Sciences of
  the United States of America}
  \bibinfo{year}{2004};\bibinfo{volume}{101}(\bibinfo{number}{14}):\bibinfo{pages}{4781--4786}.
\bibitem[{Lau et~al.(2007)Lau, Ganguli and
  Tang}]{Lau:2007:Phys-Rev-E-Stat-Nonlin-Soft-Matter-Phys:17677098}
\bibinfo{author}{Lau\xfnm[ K.]}, \bibinfo{author}{Ganguli\xfnm[ S.]},
  \bibinfo{author}{Tang\xfnm[ C.]}.
\newblock \bibinfo{title}{Function constrains network architecture and
  dynamics: a case study on the yeast cell cycle boolean network}.
\newblock \bibinfo{journal}{Phys Rev E}
  \bibinfo{year}{2007};\bibinfo{volume}{75}(\bibinfo{number}{5 Pt
  1}):\bibinfo{pages}{051907--051907}.
\bibitem[{Li et~al.(1996)Li, Helling, Tang and Wingreen}]{li1996emergence}
\newblock{*}
\bibinfo{author}{Li\xfnm[ H.]}, \bibinfo{author}{Helling\xfnm[ R.]},
  \bibinfo{author}{Tang\xfnm[ C.]}, \bibinfo{author}{Wingreen\xfnm[ N.]}.
\newblock \bibinfo{title}{Emergence of preferred structures in a simple model
  of protein folding}.
\newblock \bibinfo{journal}{Science}
  \bibinfo{year}{1996};\bibinfo{volume}{273}(\bibinfo{number}{5275}):\bibinfo{pages}{666--669}.\\
\newblock{\textbf{Revealed that entropic arguments alone, involving counting the number of sequences that lead to a fold, qualitatively explains the presence of preferred folding structures in nature.}}
\bibitem[{Wingreen et~al.(2004)Wingreen, Li and
  Tang}]{wingreen2004designability}
\bibinfo{author}{Wingreen\xfnm[ N.S.]}, \bibinfo{author}{Li\xfnm[ H.]},
  \bibinfo{author}{Tang\xfnm[ C.]}.
\newblock \bibinfo{title}{Designability and thermal stability of protein
  structures}.
\newblock \bibinfo{journal}{Polymer}
  \bibinfo{year}{2004};\bibinfo{volume}{45}(\bibinfo{number}{2}):\bibinfo{pages}{699--705}.
\bibitem[{Li et~al.(1998)Li, Tang and Wingreen}]{li1998protein}
\bibinfo{author}{Li\xfnm[ H.]}, \bibinfo{author}{Tang\xfnm[ C.]},
  \bibinfo{author}{Wingreen\xfnm[ N.S.]}.
\newblock \bibinfo{title}{Are protein folds atypical?}
\newblock \bibinfo{journal}{Proceedings of the National Academy of Sciences}
  \bibinfo{year}{1998};\bibinfo{volume}{95}(\bibinfo{number}{9}):\bibinfo{pages}{4987--4990}.
\bibitem[{Maass et~al.(2002)Maass, Natschlager and Markram}]{maass2002real}
\bibinfo{author}{Maass\xfnm[ W.]}, \bibinfo{author}{Natschlager\xfnm[ T.]},
  \bibinfo{author}{Markram\xfnm[ H.]}.
\newblock \bibinfo{title}{Real-time computing without stable states: A new
  framework for neural computation based on perturbations}.
\newblock \bibinfo{journal}{Neural computation}
  \bibinfo{year}{2002};\bibinfo{volume}{14}(\bibinfo{number}{11}):\bibinfo{pages}{2531--2560}.
\bibitem[{Jaeger(2001)}]{Jaeger2001}
\bibinfo{author}{Jaeger\xfnm[ H.]}.
\newblock \bibinfo{title}{Short term memory in echo state networks}.
\newblock \bibinfo{journal}{GMD Report 152 German National Research Center for
  Information Technology} \bibinfo{year}{2001};.
\bibitem[{White et~al.(2004)White, Lee and Sompolinsky}]{white2004short}
\bibinfo{author}{White\xfnm[ O.]}, \bibinfo{author}{Lee\xfnm[ D.]},
  \bibinfo{author}{Sompolinsky\xfnm[ H.]}.
\newblock \bibinfo{title}{Short-term memory in orthogonal neural networks}.
\newblock \bibinfo{journal}{Phys Rev Lett}
  \bibinfo{year}{2004};\bibinfo{volume}{92}(\bibinfo{number}{14}):\bibinfo{pages}{148102}.
\bibitem[{Ganguli et~al.(2008)Ganguli, Huh and Sompolinsky}]{ganguli2008memory}
\bibinfo{author}{Ganguli\xfnm[ S.]}, \bibinfo{author}{Huh\xfnm[ D.]},
  \bibinfo{author}{Sompolinsky\xfnm[ H.]}.
\newblock \bibinfo{title}{{Memory traces in dynamical systems}}.
\newblock \bibinfo{journal}{Proc Natl Acad Sci}
  \bibinfo{year}{2008};\bibinfo{volume}{105}(\bibinfo{number}{48}):\bibinfo{pages}{18970}.
\bibitem[{Ganguli and Sompolinsky(2010{\natexlab{b}})}]{ganguli2010short}
\bibinfo{author}{Ganguli\xfnm[ S.]}, \bibinfo{author}{Sompolinsky\xfnm[ H.]}.
\newblock \bibinfo{title}{Short-term memory in neuronal networks through
  dynamical compressed sensing}.
\newblock In: \bibinfo{booktitle}{Neural Information Processing Systems
  (NIPS)}. \bibinfo{year}{2010}{\natexlab{b}},.
\bibitem[{Fusi et~al.(2005)Fusi, Drew and Abbott}]{fusi2005cascade}
\bibinfo{author}{Fusi\xfnm[ S.]}, \bibinfo{author}{Drew\xfnm[ P.J.]},
  \bibinfo{author}{Abbott\xfnm[ L.F.]}.
\newblock \bibinfo{title}{{{C}ascade models of synaptically stored memories}}.
\newblock \bibinfo{journal}{Neuron}
  \bibinfo{year}{2005};\bibinfo{volume}{45}(\bibinfo{number}{4}):\bibinfo{pages}{599--611}.
\newblock \doi{\bibinfo{doi}{10.1016/j.neuron.2005.02.001}}.
\bibitem[{Lahiri and Ganguli(2014)}]{ganguli14memfrontier}
\bibinfo{author}{Lahiri\xfnm[ S.]}, \bibinfo{author}{Ganguli\xfnm[ S.]}.
\newblock \bibinfo{title}{A memory frontier for complex synapses}.
\newblock In: \bibinfo{booktitle}{Neural Information Processing Systems
  (NIPS)}. \bibinfo{year}{2014},.
\bibitem[{Prinz et~al.(2004)Prinz, Bucher and Marder}]{prinz2004similar}
\bibinfo{author}{Prinz\xfnm[ A.A.]}, \bibinfo{author}{Bucher\xfnm[ D.]},
  \bibinfo{author}{Marder\xfnm[ E.]}.
\newblock \bibinfo{title}{Similar network activity from disparate circuit
  parameters}.
\newblock \bibinfo{journal}{Nature neuroscience}
  \bibinfo{year}{2004};\bibinfo{volume}{7}(\bibinfo{number}{12}):\bibinfo{pages}{1345--1352}.
\bibitem[{Schulz et~al.(2007)Schulz, Goaillard and
  Marder}]{schulz2007quantitative}
\bibinfo{author}{Schulz\xfnm[ D.J.]}, \bibinfo{author}{Goaillard\xfnm[ J.M.]},
  \bibinfo{author}{Marder\xfnm[ E.E.]}.
\newblock \bibinfo{title}{Quantitative expression profiling of identified
  neurons reveals cell-specific constraints on highly variable levels of gene
  expression}.
\newblock \bibinfo{journal}{Proceedings of the National Academy of Sciences}
  \bibinfo{year}{2007};\bibinfo{volume}{104}(\bibinfo{number}{32}):\bibinfo{pages}{13187--13191}.
\bibitem[{O'Leary et~al.(2013)O'Leary, Williams, Caplan and
  Marder}]{o2013correlations}
\bibinfo{author}{O'Leary\xfnm[ T.]}, \bibinfo{author}{Williams\xfnm[ A.H.]},
  \bibinfo{author}{Caplan\xfnm[ J.S.]}, \bibinfo{author}{Marder\xfnm[ E.]}.
\newblock \bibinfo{title}{Correlations in ion channel expression emerge from
  homeostatic tuning rules}.
\newblock \bibinfo{journal}{Proceedings of the National Academy of Sciences}
  \bibinfo{year}{2013};\bibinfo{volume}{110}(\bibinfo{number}{28}):\bibinfo{pages}{E2645--E2654}.
\bibitem[{Shakespeare(2001)}]{shakespeare2001tempest}
\bibinfo{author}{Shakespeare\xfnm[ W.]}.
\newblock \bibinfo{title}{The tempest}; vol.~\bibinfo{volume}{9}.
\newblock \bibinfo{publisher}{Classic Books Company}; \bibinfo{year}{2001}.

\end{thebibliography}







\end{document}